\documentclass[12pt,a4paper]{article}

\usepackage{graphicx}
\usepackage{amsmath} 
\usepackage{amsfonts} 
\usepackage{multirow}
\usepackage{natbib}
\usepackage{color}
\usepackage{enumitem}
\usepackage[hidelinks]{hyperref}
\usepackage[title]{appendix}

\def\R{\mathbb{R}}
\def\N{\mathbb{N}}

\def\argmax{\mathop{\rm arg\,max}\limits}

\usepackage{amssymb}
\usepackage{amsthm}
\theoremstyle{plain}

\newtheorem{theorem}{Theorem}

\theoremstyle{remark}

\usepackage[affil-sl]{authblk}

\begin{document}

\title{Inference for the proportional odds cumulative logit model with monotonicity constraints for ordinal predictors and ordinal response}

%\author[1]{Javier Espinosa-Brito}
%\author[2]{Christian Hennig}
%\affil[1]{University of Santiago of Chile (USACH), Faculty of Administration and Economics, Economics Department, Chile.}
%\affil[2]{Universita di Bologna, Department of Statistical Sciences ``Paolo Fortunati'', Italy.}

\author[1]{Javier Espinosa-Brito}
\author[2]{Christian Hennig}
\affil[1]{University of Santiago of Chile (USACH), Faculty of Administration and Economics, Economics Department, Chile.}
\affil[2]{Universita di Bologna, Department of Statistical Sciences ``Paolo Fortunati'', Italy.}

\date{}

\maketitle

\begin{abstract}
The proportional odds cumulative logit model (POCLM) is a standard regression model for an ordinal response. In this article, ordinality of predictors is incorporated by imposing monotonicity constraints on their corresponding parameters. It is shown that the parameter estimates of an unconstrained model are asymptotically equivalent to the ones of a constrained model when they are in the interior set of the parameter space. This is used in order to derive asymptotic confidence regions and tests for the constrained model based on maximum likelihood estimation, involving simple modifications for finite samples. The finite sample coverage probability of the confidence regions is investigated by simulation. Tests concern the effect of individual variables, monotonicity, and a specified monotonicity direction. The power of the proposed test on the effect of an ordinal predictor is compared with inference based on one-dimensional confidence intervals via simulations. The methodology is applied on real data related to the assessment of school performance.


\end{abstract}

\section{Introduction} \label{sec:IntroductionCh3}

Regression problems are treated in which the response variable is ordinal and one or more of the predictors are ordinal as well, potentially in the presence of additional interval-scaled or categorical predictors. When considering regression relations between variables, monotonicity of the parameters associated to categories has a special status regarding ordinal variables. The ordinal scale type implies that the ordering of the possible values is meaningful, whereas there is no meaningful quantitative distance between them. 
A monotonic relationship between an ordinal predictor and an ordinal response means that changes in the predictor affect the response ``consistently'', i.e., whenever the predictor increases, the response tends to move in the same direction (be it larger or smaller). Arguably, monotonicity for ordinal variables is analogous to linearity for interval scaled measurements with meaningful quantitative differences, where changing the explanatory variable by a certain amount will on average change the response consistently, i.e., by an amount that is everywhere the same. In this sense the monotonic relationship between ordinal variables in a regression is a key reference point, as is the linearity assumption between interval scaled variables. Although of course not all relations between ordinal variables are monotonic, monotonicity can be seen as default, unless contradicted by the data or specific background knowledge.

We treat a maximum likelihood estimator (MLE) in a proportional odds cumulative logit model (POCLM, \cite{mccullagh1980regression}), in which relationships between ordinal variables are constrained to be monotonic, i.e., for an isotonic association the restriction on the $p_s$ parameters of the ordinal explanatory variable $s$ is
$\beta_{s,1}=0 \leq \beta_{s,2}\leq\cdots\leq\beta_{s,p_s}$,
whereas if the constraints are antitonic, then
$\beta_{s,1}=0 \geq \beta_{s,2}\geq\cdots\geq\beta_{s,p_s}$. Such constraints have been proposed by \cite{espinosa2019constrained}. We pay special attention to the relation between the constrained and the unconstrained MLE. We consider confidence regions (CRs) and monotonicity direction detection. In Section \ref{AICM_sec:tests}, we use the duality between tests and  CRs in order to test whether (i) there is no impact of an ordinal predictor, (ii) its parameters are monotonic, and (iii) there is a specific monotonicity direction for the impact of the ordinal predictor (isotonic or antotonic).

\cite{espinosa2019constrained} proposed to decide about the appropriateness of the monotonicity assumption and monotonicity directions based on one-dimensional confidence intervals (CIs) for parameters, which uses less information than multivariate CRs taking into account dependence between parameters, particularly referring to the same predictor, as proposed here. The main idea of the present approach does not exclusively apply to the POCLM and can also be used with other models, but we will focus on the POCLM.

There is a lot of work on constrained inference with some results concerning monotonicity constraints as a special case of linear constraints, see, e.g., \cite{Shapiro88,dupacova1988optim,SilSen05}. 
In particular, in many settings based (potentially asymptotically) on Gaussian distributions, the distribution of the loglikelihood ratio test statistic involving linearly constrained MLEs is a mixture of $\chi^2$-distributions with degrees of freedom smaller or equal to the $\chi^2$-distribution that applies in the corresponding unconstrained case (\cite{GoHoMo82}).

The association among categorical variables, with some of them being ordinal, has been addressed by analysing multiway tables, where the conditional distribution of a subset of variables given the remaining ones and testing hypotheses involving equality and inequality constraints may be of interest (\cite{colombi2001marginal}, \cite[Chapter 6]{agresti2010analysis}). Extensions of RC (row/column) models have also been proposed allowing order-restricted inference when analysing contingency tables (\cite{bartolucci2002extended}). These methods aren't direct competitors of our approach, as we treat regression here, allowing for potentially non-ordinal variables.

Additional complexity arises because (a) we allow for both isotonic and antitonic relationships, so that the constrained parameter space is not convex, and (b) the distribution of the constrained loglikelihood ratio statistic for one ordinal predictor of interest will depend on the unknown parameter values for other ordinal or even non-ordinal predictors if present, for which we allow. Standard results for constrained inference have not yet been adapted to our situation, although this would be a promising direction for future research.
\cite{SelLia87} and \cite{KopSin11} demonstrate that constraints in nuisance parameters can require more sophisticated distributions of the test statistic than  mixtures of $\chi^2$-distributions. 

Further work regarding constrained inference problems, potentially with nuisance parameters, contains ideas that may be applied to the present problem (\cite{Ketz18,AMVZCaGo20,CoxShi22,ChNeSa23,kitamura2004empirical,stoye2009more,andrews2010inference,bennett2013inference}), but these ideas have not been elaborated yet for the problem treated here to our knowledge. We are not aware of methodology that can be directly applied to our setting, or that treats a parameter space of the specific shape resulting from allowing both isotonicity and antitonicity. In situations where a single monotonicity direction can be assumed for all involved ordinal predictors, based on subject matter knowledge, the issue of non-convexity is not present, and deriving theory for parameters on the boundary may not be so difficult. We believe, however, that the monotonicity direction can often not be assumed to be known. 

If there are several explanatory variables in a standard linear regression model, it is well known that the sign of the coefficient estimate belonging to a particular numerical predictor can turn out to be the opposite of what was a priori assumed, based on background knowledge or exploratory analysis, due to the dependence structure between predictors. The p-values for standard t-tests given out by software for fitting linear regression are two-sided, not requiring to fix the linearity direction. Note also that it is invalid to decide the monotonicity direction to be tested based on exploratory analysis on the same data, as standard test theory assumes that tests are not chosen conditionally on the data. 

Here is an example why one cannot necessarily specify the monotonicity direction to be tested against in advance. We have worked on a confidential data set in which an overall ordinal rating of a restaurant was modelled as dependent on ratings of several aspects of the restaurant experience. People were asked regarding their most recent restaurant experience, meaning that there were different restaurants involved. All Spearman correlations between the variables turned out to be significantly positive,  indicating that generally positive or generally negative ratings on (almost) all aspects were quite common. Running a multiple ordinal regression, however, the impact of the variable ``neatness of waiters'' for predicting overall satisfaction turned out to have a significantly negative monotonicity direction - higher overall ratings were predicted from lower ratings of ``neatness of waiters'' taking all other variables into account. This was a surprise. Before seeing the data nobody would have specified this monotonicity direction, and testing monotonicity would have been needed to be two-sided. Looking into the data in more detail, the result can be explained by the fact that a certain group of participants rated the neatness of waiters low but the overall experience higher than predicted from the other variables, probably because they prefer a nonconformist style of clothing. For other participants, any positive connection between the rating of neatness of waiters and the overall experience could apparently be explained by the overall correlation pattern. 

Turning back to methodology, bootstrap may in principle be available, but can be inconsistent in such a situation (\cite{Andrews00}). \cite{CoKeSa00} treats a further issue with constrained likelihood inference. 
An alternative approach for regression with ordinal predictors would be to represent them by latent variables (\cite{moustaki2000latent}) or optimal scaling (\cite{de2009gifi}). Work on isotonic regression (\cite{stout2015isotonic}) is also related to the present approach. 

As the adaptation of existing approaches for constrained inference to the present setup is not straightforward, we take a different direction. Whereas much of the constrained inference literature focuses on tests, the starting point for the inference presented here is CRs, and we explore what can be done based on the standard asymptotics for unconstrained inference. 
Tests can then be constructed by the equivalence of tests and CRs. It has been noted in the literature (e.g., \cite{Shapiro88,KopSin11}) that in many cases the standard $\chi^2$-asymptotics apply to parameters in the interior of the parameter space, whereas the cited theory of constrained inference focuses on parameters on the boundary, as is required for computing p-values under ``worst case'' conditions in the null hypothesis. Focusing on CRs means that ``most'' of the considered parameter values are in the  interior, and boundary behaviour is rather exceptional. We do not deny that the behaviour on the boundary of the parameter space is relevant. Sometimes it is of interest to test a boundary null hypothesis. Also, for finite samples, boundary effects are relevant if the true parameter is close to the boundary but not necessarily on it. 

Coverage probabilities of CRs ignoring constraints 
are still valid for boundary parameters. The problem with these CRs is (a) their overlap with the constrained parameter set may be empty or very small in case that the data are to some extent in conflict with the constraints, and (b) that otherwise they can be too big, i.e., associated tests are too conservative and less powerful than they could be if based on distribution theory that take the constraints properly into account (as illustrated in the Introduction of \cite{SilSen05}).
Without having such a theory available, we investigate in Section \ref{AICM_sec:CRsAndCovProb} how well modifying unconstrained inference by just using the constrained estimator works. Given that all inference is based on asymptotic theory, one result is that coverage probabilities can be slightly lower than nominal for finite samples, particularly small ones. This suggests that the higher power to be potentially achieved by proper boundary theory, which here can only be asymptotic, may come at the price of even worse coverage and anticonservativity of tests in finite samples, as the resulting CRs may be smaller.

Section \ref{sec:AsymptUMLE} introduces the unconstrained POCLM and reviews its inference and asymptotic theory. Not only are CRs for the unconstrained POCLM a basis for the CRs for the constrained POCLM, they are of interest in their own right when making statements about whether the data are compatible with none, one, or both possible monotonicity directions for the ordinal predictors. Section \ref{AICM_sec:MonotConstraints} introduces the monotonicity constraints for ordinal predictors. It is possible to leave the monotonicity direction unspecified, or to constrain the parameters belonging to a variable to be isotonic, or antitonic. In order to use the unconstrained asymptotic theory for the constrained model, a general theorem is shown that constrained estimators defined by optimization are asymptotically equivalent to their unconstrained counterparts on the interior of the constrained parameter space.
This result is used in Section \ref{AICM_sec:AsympCRs} in order to define confidence regions and tests for the constrained POCLM. Confidence regions based on unconstrained asymptotics may in finite samples contain parameters that do not fulfill monotonicity constraints, and adaptations for this case are proposed. Coverage probabilities and test performances are investigated in a simulation study in Section \ref{AICM_sec:CRsAndCovProb}. The methodology is applied to a data set on school performance in Chile in Section \ref{sec:realData}. Section \ref{sec:Conclusions} concludes the paper.

\section{The unconstrained POCLM} \label{sec:AsymptUMLE}

Consider a regression with an ordinal response $z_i,\ i=1,\ldots,n,$ $n$ being the number of independent observations, with $k$ categories. There are $t$ ordinal predictors (OPs; $x_{i,s,h_s}$ refers to a dummy variable for ordinal predictor $s=1,\ldots,t$ and category $h_s=1,\ldots,p_s$) and $v$ non-ordinal predictors $x_{n,u},\ u=1,\ldots,v$, which can be quantitative variables and/or dummy variables representing categories of a nominal predictor. According to the POCLM (\cite{mccullagh1980regression}), 
\begin{align}
\begin{split}
\text{logit}[P(z_{i}\leq j | \mathbf{x}_i)]=&\alpha_j+\sum_{s=1}^{t}\sum_{h_s=2}^{p_s}\beta_{s,h_s}x_{i,s,h_s} +\sum_{u=1}^{v}\beta_{u}x_{i,u},\label{eq:Model_eq}\\
j=&1,\ldots, k-1, 
\end{split}
\end{align}
with \begin{align}
-\infty < \alpha_1 < \cdots < \alpha_{k-1} < \infty, \label{AICM_eq:POCLM_OrderedParam_altogether}
\end{align} and where the first category for each predictor is the baseline category, i.e., $\beta_{s,1}=0$. In addition, 
$$\text{logit}[P(z_{i}\leq j | \mathbf{x}_i)]=\text{log}\bigg{[}\frac{P(z_{i}\leq j | \mathbf{x}_i)}{1-P(z_{i}\leq j | \mathbf{x}_i)}\bigg{]}.$$

The dimensionality of the parameter space is $p=(k-1)+\sum_{s=1}^t (p_s-1)+v$, defining the $p$-dimensional parameter vector as
\begin{align}
\boldsymbol{\gamma}'&=(\boldsymbol{\alpha}',\boldsymbol{\beta}') \nonumber \\
&=(\boldsymbol{\alpha}',\boldsymbol{\beta}_{(ord)}',\boldsymbol{\beta}_{(nonord)}') \nonumber \\
&=(\boldsymbol{\alpha}',\boldsymbol{\beta}_1',\ldots,\boldsymbol{\beta}_t',\beta_1,\ldots,\beta_v), \nonumber 
\end{align}
where $\boldsymbol{\beta}_{(ord)}$ is the parameter vector associated with the ordinal predictors, and $\boldsymbol{\beta}_{(nonord)}$ is the one associated with the non-ordinal predictors.

With $\mathbf{z}=(z_1,\ldots,z_n),\ \mathbf{X}_n=(\mathbf{x}_1,\ldots,\mathbf{x}_n)$, the log-likelihood function is
\begin{align}
\ell(\boldsymbol{\gamma})=\sum_{i=1}^n\sum_{j=1}^k{y_{ij}}\log\pi_j(\mathbf{x}_i), \label{eq:CRMT_likelihood_Model}
\end{align}
where $y_{ij}=1(z_i=j)$, $1(\bullet)$ denotes the indicator function,
and $\pi_j(\mathbf{x}_i)=P(z_{i}= j | \mathbf{x}_i)$, i.e.,
\begin{align}
&\pi_j(\mathbf{x}_i)=\frac{e^{\alpha_j+\sum_{s=1}^{t}\sum_{h_s=2}^{p_s}\beta_{s,h_s}x_{i,s,h_s}+\sum_{u=1}^{v}\beta_{u}x_{i,u}}}{1+e^{\alpha_j+\sum_{s=1}^{t}\sum_{h_s=2}^{p_s}\beta_{s,h_s}x_{i,s,h_s}+\sum_{u=1}^{v}\beta_{u}x_{i,u}}} \nonumber \\
&\quad \quad \quad -\frac{e^{\alpha_{j-1}+\sum_{s=1}^{t}\sum_{h_s=2}^{p_s}\beta_{s,h_s}x_{i,s,h_s}+\sum_{u=1}^{v}\beta_{u}x_{i,u}}}{1+e^{\alpha_{j-1}+\sum_{s=1}^{t}\sum_{h_s=2}^{p_s}\beta_{s,h_s}x_{i,s,h_s}+\sum_{u=1}^{v}\beta_{u}x_{i,u}}}. \label{AICM_eq:likelihood_pi}
\end{align}
The parameter space for model \eqref{eq:Model_eq} is 
\begin{align}
U_{U}=\{ &\boldsymbol{\gamma}'=(\boldsymbol{\alpha}',\boldsymbol{\beta}_{1}',\ldots,\boldsymbol{\beta}_{t}',\boldsymbol{\beta}_{(nonord)}') \in \mathbb{R}^{p} : -\infty < \alpha_1 < \cdots < \alpha_{k-1} < \infty \}, \label{AICM_eq:UnconstrainedSpaceOP_altogether}
\end{align}
where $p=k-1+\sum_{s=1}^{t}(p_s-1)+v$. We refer to this as the ``unconstrained model'' (thus ``$U_{U}$'') despite the constraints on the 
$\alpha$-parameters, because in Section \ref{AICM_sec:MonotConstraints} 
constraints will be introduced for the $\boldsymbol{\beta}$-vectors of ordinal predictors.

The unconstrained maximum likelihood estimator (UMLE) is 
\begin{align}
\hat{\boldsymbol{\gamma}} = \argmax_{\boldsymbol{\gamma} \in U_{U}} \ell(\boldsymbol{\gamma}), \label{eq:CRMT_unconstrainedMaxLikelihood}
\end{align}
then $\hat{\boldsymbol{\gamma}}$ is the vector of UMLEs belonging to the parameter space $U_{U}$.

The score function and the Fisher information matrix of the first $n$ observations are
\begin{align}
\mathbf{s}_n(\boldsymbol{\gamma}) &= \partial \log L(\boldsymbol{\gamma}|\mathbf{z}_n,\mathbf{X}_n) / \partial \boldsymbol{\gamma}, \label{AICM:eq_ScoreFunction} \\
\mathbf{F}_n(\boldsymbol{\gamma}) &= \text{cov} (\mathbf{s}_n(\boldsymbol{\gamma})). \label{AICM:eq_FisherInfoMatrix}
\end{align}

\subsection{Asymptotic results for the unconstrained POCLM}
\label{sec:UPOCLM_Theory}

Asymptotic theory for generalized linear models (GLMs), including the unconstrained POCLM, is given in \cite{fahrmeir1985consistency,fahrmeir1986asymptotic,fahrmeir1987chisq}.

Under conditions given in \cite{fahrmeir1986asymptotic}, the probability converges to one that a unique MLE $\hat{\boldsymbol{\gamma}}$ exists. Any sequence of MLEs $(\hat{\boldsymbol{\gamma}}_n)_{n\in\N}$ is weakly consistent and asymptotically normal,
\begin{align}
(\mathbf{F}_n(\boldsymbol{\gamma})^{-1/2})'(\hat{\boldsymbol{\gamma}}_n-\boldsymbol{\gamma}) \stackrel{\mathcal{L}}{\to} \mathcal{N}_p(\mathbf{0},\mathbf{I}_p), \label{tasymnormal}
\end{align}
where ``$\stackrel{\mathcal{L}}{\to}$'' denotes convergence in distribution, $\mathcal{N}_p(\mathbf{0},\mathbf{I}_p)$ is the $p$-variate Gaussian distribution with mean vector $\mathbf{0}$ and unit matrix $\mathbf{I}_p$ as covariance matrix, and $ \mathbf{F}_n(\boldsymbol{\gamma})^{-1/2}$ is the Cholesky square root of $\mathbf{F}_n(\boldsymbol{\gamma})$.

Consider a standard loglikelihood ratio test for the linear hypothesis 
\begin{align}
H_0:\ \mathbf{C}\boldsymbol{\gamma}=\boldsymbol{\xi}\mbox{ against }H_1:\ 
\mathbf{C}\boldsymbol{\gamma}\neq\boldsymbol{\xi},
\label{eq:linearhyp}
\end{align}
where $\mathbf{C}$ is an $r\times p$-matrix of full row rank $r$, $\boldsymbol{\xi}\in\R^r$, and $H_0$ has a nonempty intersection with the parameter space. Let
\begin{align} \label{eq:likrat}
R_n=2[\ell(\hat{\boldsymbol{\gamma}}) - \ell(\tilde{\boldsymbol{\gamma}})]
\end{align}
be the standard loglikelihood ratio statistic, where $\tilde{\boldsymbol{\gamma}}$ is the MLE within $H_0$, fulfilling $\mathbf{C}\tilde{\boldsymbol{\gamma}}=\boldsymbol{\xi}$, and  
\begin{align}
W_n=\left(\mathbf{C}\hat{\boldsymbol{\gamma}}-\boldsymbol{\xi}\right)'\left[\mathbf{C}\mathbf{F}_n^{-1}(\hat{\boldsymbol{\gamma}})\mathbf{C}'\right]^{-1}\left(\mathbf{C}\hat{\boldsymbol{\gamma}}-\boldsymbol{\xi}\right)
\end{align}
be the Wald statistics.

\cite{fahrmeir1986asymptotic} give conditions so that, if $H_0$ in \eqref{eq:linearhyp} is true, then $R_n$ and 
$W_n$ are asymptotically equivalent, and
\begin{align}
R_n,\ W_n \stackrel{\mathcal{L}}{\to} \chi^2_r. \label{tlr}
\end{align}

We will mainly consider asymptotic CRs based on $R_n$ here. One-dimensional CIs for a parameter $\beta$ of level $1-\nu$ (as used in \cite{espinosa2019constrained}) are usually constructed as $\hat{\beta} \pm z_{1-\nu/2}(SE_{\hat{\beta}}),$ where $z_{1-\nu/2}$ is the ${1-\nu/2}$-quantile of the Gaussian distribution, and the estimated standard error $SE_{\hat{\beta}}$ can be obtained
from the asymptotic covariance matrix. In \eqref{tasymnormal}, $\mathbf{F}_n(\boldsymbol{\gamma})$ depends on the true parameter $\boldsymbol{\gamma}$, which in reality is unknown. In order to apply \eqref{tasymnormal}, $\mathbf{F}_n(\boldsymbol{\gamma})$ needs to be replaced by  $\mathbf{F}_n(\hat{\boldsymbol{\gamma}})$. Extending \eqref{tasymnormal} to this case requires additional conditions, see \cite{fahrmeir1985consistency}. However, the resulting CI is equivalent to the one that can be obtained from the Wald statistic $W_n$ with $r=1$, and therefore it is valid according to \eqref{tlr}. We normally prefer the loglikelihood ratio statistic $R_n$ to $W_n$ for inference, see \cite{harrell2001rms} for a discussion.

\section{Monotonicity constraints for the POCLM}\label{AICM_sec:MonotConstraints}
\subsection{The constrained parameter space}\label{sec:ParamSpaceCMLE}
When ordinal predictors are treated as if they were
of nominal scale type, no monotonicity constraints are imposed on any of the $t$ vectors $\boldsymbol{\beta}_s$, $s=1,\ldots,t$, and the parameter space is $U_{U}$.

In order to take their ordinality into account,
\cite{espinosa2019constrained} proposed monotonic constraints on the $p_s-1$ parameters associated with the categories of an ordinal predictor $s$. Monotonicity is allowed to be either isotonic or antitonic.
The isotonic constraints are
\begin{align}
0 \leq \beta_{s,2}\leq\cdots\leq\beta_{s,p_s}, \quad \forall s \in \mathcal{I}, \label{AICM_eq:NonStrictIsotonicConstraints_altogether}
\end{align}
where $\mathcal{I}\subseteq \mathcal{S}$, with $\mathcal{S}=\{1,2,\ldots,t\}$, and the antitonic constraints are
\begin{align}
0 \geq \beta_{s,2}\geq\cdots\geq\beta_{s,p_s}, \quad \forall s \in \mathcal{A}, \label{AICM_eq:NonStrictAntitonicConstraints_altogether}
\end{align}
where $\mathcal{A}\subseteq \mathcal{S}$.

The model under monotonicity constraints, i.e., requiring $s\in \mathcal{A}\cup\mathcal{I}$ for all ordinal variables $s$ in \eqref{eq:Model_eq} will be referred to as the constrained model. Even if a variable is in fact ordinal, a researcher may have reasons to not enforce constraints, in which case it can be treated as non-ordinal in the model.

$U_{C}$ is the parameter set that constrains all ordinal variables to fulfill monotonicity constraints but allows for both monotonicity directions:
\begin{align}
U_{C}=\{ \boldsymbol{\gamma}'=(\boldsymbol{\alpha}',& \boldsymbol{\beta}_{1}',\ldots,\boldsymbol{\beta}_{t}',\boldsymbol{\beta}_{(nonord)}') \in \R^{p} : -\infty < \alpha_1 < \cdots < \alpha_{k-1} < \infty, \nonumber \\
&[(\beta_{s,2}\ge 0,\beta_{s,h_s}\ge\beta_{s,h_s-1}) \text{ or } (\beta_{s,2}\le 0,\beta_{s,h_s}\le\beta_{s,h_s-1})] \label{AICM_eq:MonotonicAllSpace_altogether} \\ 
&\forall (s,h_s) \in \mathcal{S} \times \{3,\ldots,p_s\} \}. \nonumber
\end{align}
$U_{D}$ is also a constrained parameter set, but it fixes the monotonicity directions for each ordinal variable: 
% In this sense, $U_{D}$ is direction constrained:
\begin{align}
U_{D}=\{ \boldsymbol{\gamma}'=(\boldsymbol{\alpha}',&\boldsymbol{\beta}_{1}',\ldots,\boldsymbol{\beta}_{t}',\boldsymbol{\beta}_{(nonord)}') \in \R^{p} : -\infty < \alpha_1 < \cdots < \alpha_{k-1} < \infty, \nonumber \\
&(\beta_{s,2}\ge 0 , \beta_{s,h_s}\ge \beta_{s,h_s-1}) \quad \forall (s,h_s) \in \mathcal{I} \times \{3,\ldots,p_s\}, \label{AICM_eq:MonotonicSpace_altogether} \\
&(\beta_{s,2}\le 0 , \beta_{s,h_s}\le \beta_{s,h_s-1}) \quad \forall (s,h_s) \in \mathcal{A} \times \{3,\ldots,p_s\} \}. \nonumber
\end{align}
Whereas estimating in $U_D$ is similar to many constrained inference problems in the literature, we are not aware of treatments of a set of constraints like $U_C$, a non-convex union of cones. 

In practice equality for the $\beta$-parameters should be allowed, because in case the unconstrained optimizer of the likelihood violates monotonicity constraints, the constrained optimizer will only exist if equality is allowed. This is different from equality for the $\alpha$-parameters, because equality of subsequent $\alpha$ parameters would imply that there is a category $j\in\{1,\ldots,k\}$ with $P(z_i=j)=0,\ i-1,\ldots,n$, in which case this category can be dropped from the model.

The constrained maximum likelihood estimator (CMLE) is 
\begin{align}
\hat{\boldsymbol{\gamma}}_C = \argmax_{\boldsymbol{\gamma} \in U_{C}} \ell(\boldsymbol{\gamma}). \label{eq:CRMT_constrainedMaxLikelihood}
\end{align} 
For computation of $\hat{\boldsymbol{\gamma}}_C$ see \cite{espinosa2019constrained}.

There may also be interest in direction constrained MLEs (DMLEs) with fully pre-specified monotonicity directions with $\mathcal{I}$ and $\mathcal{A}$ chosen suitably,
\begin{align}
\hat{\boldsymbol{\gamma}}_D = \argmax_{\boldsymbol{\gamma} \in U_{D}} \ell(\boldsymbol{\gamma}), \label{eq:CRMT_directionMaxLikelihood}
\end{align}
or even in partially direction constrained MLEs (PMLEs) that optimize $\boldsymbol{\gamma}$ over a parameter space with some monotonicity directions fixed and others left open. These estimators can be computed using the R-function \verb|maxLik| from the package of the same name (\cite{maxLik2011}).

\subsection{Asymptotics for constrained estimators}
The asymptotic theory for the UMLE presented above makes sure that for large enough $n$ it is arbitrarily close to the true parameter. In case that monotonicity constraints indeed hold, the UMLE can therefore be expected to be in $U_{D}$ with arbitrarily large probability as long as the true parameter vector is in the set $U_{D}^o$, where for a set $S\subseteq\R^p$, $S^o$ denotes the
interior set. %, i.e., 
%$$
%S^o=\{x:\ \exists \epsilon>0: \|x-y\|<\epsilon\Rightarrow y\in S\}.
%$$

$U_{D}^o$ is $U_{D}$ with all ``$\ge$'' and ``$\le$'' replaced by ``$>$'' and ``$<$'', respectively. In this case, the UMLE and the CMLE will be equal with probability approaching one, and asymptotically equivalent. 

We will state this rather obvious fact as a general theorem for constrained estimators, which we have not been able to find in the literature; \cite{liew1976constrained} showed a corresponding result for the least squares estimator; 
\cite{SilSen05,Shapiro88,KopSin11} mention the applicability of standard theory in the interior of the parameter space in various situations. 

For observations $x_i,\ i\in \N$ in some space $ \mathcal{X}$, consider estimators $(T_n)_{n\in\N}$ of parameters $\theta$ from some parameter set $\Theta\subseteq \R^p$ that are defined by optimization of a function $h_n$:
\begin{align}
T_n(x_1,\ldots,x_n)=\argmax_{\theta\in\Theta}h_n(x_1,\ldots,x_n;\theta).
\end{align}
In case the argmax does not exist or is not unique, $T_n$ can be defined as taking any value.
The issue of interest here is whether the following properties hold for
a constrained estimator assuming that they hold for the unconstrained $T_n$:
\begin{enumerate}
\item[(C1)] The probability that the argmax in the definition of $T_n$ exists and is unique converges to 1 for $n\to\infty$.
\item[(C2)] $T_n$ is weakly consistent for $\theta$ (the results below also hold if ``weakly'' is replaced by ``strongly'').
\item[(C3)] For given functions $G_n$ and distribution $Q$, assuming $\theta_0\in \Theta$ as the true parameter,
$$
G_n(T_n;\theta_0)\stackrel{\mathcal{L}}{\to} Q.
$$     
\end{enumerate}
Constrained estimation is defined by assuming 
$$
\theta\in\tilde{\Theta}\subset\Theta,\ \tilde{T}_n(x_1,\ldots,x_n)=\argmax_{\theta\in\tilde{\Theta}}h_n(x_1,\ldots,x_n;\theta).
$$
If the argmax in the definition of $T_n$ exists uniquely, and $T_n\in\tilde{\Theta}$, then $T_n=\tilde{T}_n$. 
\begin{theorem}\label{tconstrain} With the notation introduced above, assuming that
the true $\theta_0\in\tilde{\Theta}^o$,
\begin{enumerate}
\item[(a)] If $T_n$ fulfills (C1) and (C2), then $\tilde{T}_n$ fulfills (C1) and (C2), and 
$$
\lim_{n\to\infty} P\{T_n=\tilde{T}_n\}=1.
$$
\item[(b)] If $T_n$ fulfills (C1), (C2), and (C3), then 
$\tilde{T}_n$ also fulfills (C3).
\end{enumerate}
\end{theorem}
\begin{proof}
(C2) for $T_n$ means that 
for all $\epsilon>0:\ P\{\|T_n-\theta_0\|<\epsilon\}\to 1$. 
Consider $\theta_0\in \tilde{\Theta}^o$ and $\epsilon>0$ so that
$$ 
U_\epsilon=\{\theta\in\Theta:\ \|\theta-\theta_0\|<\epsilon\}\subseteq \tilde{\Theta}.
$$
If the argmax exists and is unique (which happens with $P\to 1$ because of (C1)), if $T_n\in U_\epsilon$, then $\tilde{T}_n=T_n$, therefore 
$$
P\{T_n=\tilde{T}_n\}\to 1 \mbox{ and }
P\{\|\tilde{T}_n-\theta_0\|<\epsilon\}\to 1.
$$ 
This holds for all $\epsilon>0$, as $U_\epsilon\subseteq \tilde{\Theta}$ for arbitrarily small $\epsilon$, and if $\epsilon^*$ is so large that $U_{\epsilon^*}\not\subseteq \tilde{\Theta}$, then $\epsilon^*>\epsilon$ and 
$$
P\{\|\tilde{T}_n-\theta_0\|<\epsilon^*\}\ge P\{\|\tilde{T}_n-\theta_0\|<\epsilon\}\to 1.
$$ 
Therefore (C2) holds for $\tilde{T}_n$, as does (C1), because 
$$
P\left(\{\mbox{argmax is unique and exists }\}\cap \{T_n\in U_\epsilon\subseteq \tilde{\Theta}\}\right)\to 1.
$$ 
This proves (a).

(C3) for $T_n$ means that for all continuity points $x$ of the cdf of $Q$, 
$$
P\{G_n(T_n;\theta_0)\le x\}\to Q(-\infty,x].
$$ 
Consider again $U_\epsilon\subseteq \tilde{\Theta}$. Ignoring the possibility that the argmax might not exist or be unique (the probability of which vanishes due to (C1)), let 
\begin{align}
U_{1n}=\{T_n\in U_\epsilon,\ G_n(\tilde{T}_n;\theta_0)\le x\}=\{T_n\in U_\epsilon,\ G_n(T_n;\theta_0)\le x\},\\ 
U_{2n}=\{T_n\in U_\epsilon^c,\ G_n(\tilde{T}_n;\theta_0)\le x\}. 
\end{align}
Then
$\{G_n(\tilde{T}_n;\theta_0)\le x\}=U_{1n}\cup U_{2n}$. Because of (C2), $P\{T_n\in U_\epsilon\}\to 1$, and  therefore 
$$
|P(U_{1n})-P\{G_n(T_n;\theta_0)\le x\}|\to 0,\ P(U_{2n})\to 0.
$$ 
Therefore,
\begin{align}
\lim_{n\to\infty}P\{G_n(\tilde{T}_n;\theta_0)\le x\}=\lim_{n\to\infty} P(U_{1n})=
\lim_{n\to\infty}P(G_n(T_n;\theta_0)\le x\}=Q(-\infty,x],
\end{align}
proving (C3) for $\tilde{T}_n$ and therefore (b).
\end{proof}
Theorem \ref{tconstrain} implies that \eqref{tasymnormal} and \eqref{tlr} do not only hold for the UMLE but also for the CMLE, the DMLE and PMLEs. This means that asymptotic inference for them is the same, unless $\theta_0$ is on the border of $U_{C}$, in which case the probability for the CMLE to equal the UMLE may not go to one. Weak consistency also implies that for $\theta_0\in U_{C}^o$ and $n\to\infty$ all monotonicity directions will be estimated correctly with probability converging to 1, i.e., UMLE and CMLE will be in $U_{D}$.

\section{Inference based on asymptotics}\label{AICM_sec:AsympCRs}

Simultaneous inference regarding a subset of parameters is used to build confidence regions. A case of special interest is a subset of all parameters belonging to an ordinal predictor, so that their monotonicity can be studied. In order to take into account finite samples cases where these confidence regions based on unconstrained asymptotics may contain parameters that do not fulfill monotonicity constraints, some adaptations are proposed.

Based on the equivalence between CRs and tests, the latter are proposed and discussed for three kinds of null hypotheses regarding the parameters associated to an ordinal predictor. These allow to test whether (i) there is no impact of the ordinal predictor, (ii) its parameters are monotonic, and (iii) there is a specific monotonicity direction for the impact of the ordinal predictor (isotonic or antotonic). Finally, their coverage probabilities are analysed.

\subsection{Confidence regions}\label{AICM_sec:CR}
Due to the asymptotic equivalence of UMLE and CMLE, inference about the parameters of the constrained POCLM can be based on the theory for the unconstrained POCLM, particularly \eqref{eq:likrat} and \eqref{tlr}. As opposed to \cite{espinosa2019constrained}, the results here allow for inference regarding multiple parameters. Simultaneous inference may be of particular interest regarding all parameters belonging to a variable. Unconstrained inference can be relevant for testing monotonicity. If the monotonicity assumption for ordinal variables is indeed true, one should expect higher efficiency for inference that takes the monotonicity constraints into account. Regarding constrained inference, CRs are probably most interesting; they may be interpreted as quantifying uncertainty of the ``effect size''. 

A major issue regarding Theorem \ref{tconstrain} is that its proof is based on the fact that asymptotically UMLE and CMLE are the same. CRs based on \eqref{tlr} will be, with probability converging to one, subsets of arbitrarily small neighborhoods of the MLE and therefore also of the true parameters, meaning that if the monotonicity constraints are indeed fulfilled with strict inequalities, the whole CR will only contain parameters indicating the correct monotonicity direction. 

In practice, for finite $n$, it is neither guaranteed that UMLE and CMLE are the same, nor that all parameters in such a CR indicate the same monotonicity direction, or at least fulfill any monotonicity constraint. The validity of asymptotic CRs may then be doubted.

\begin{figure}
	\centering
	\includegraphics[scale=0.57]{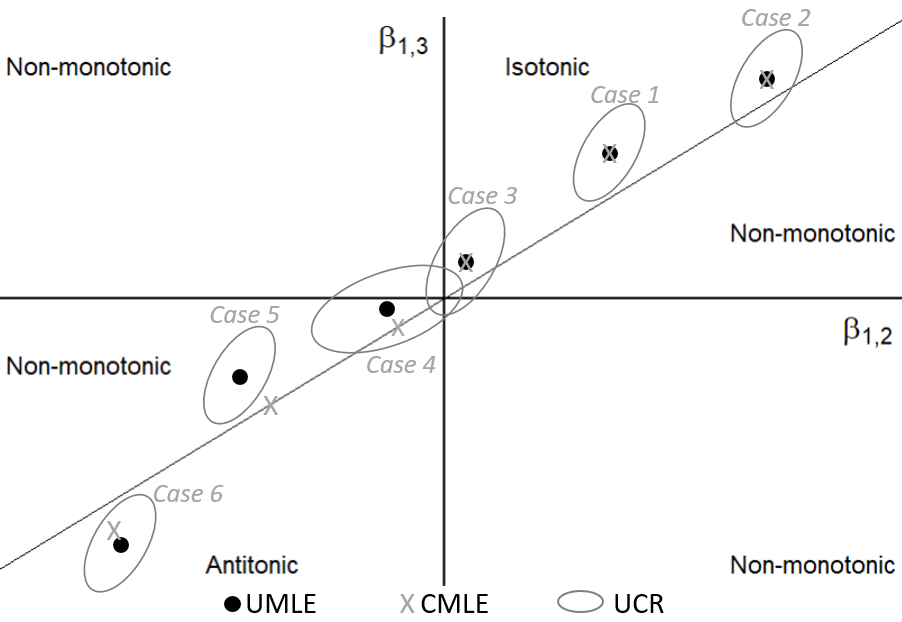}
	\caption{Illustration of some possible cases associated to the parameter vectors in UCR with equality or inequality of CMLE and UMLE.}
	\label{fig:UCRCases}
\end{figure}

For the sake of simplicity, in the following discussion, we consider CRs for a vector of parameters of interest, usually those belonging to a single ordinal variable, although ideas also apply to more general vectors fulfilling linear constraints as in \eqref{eq:linearhyp}. For a vector with $r$ parameters of interest, $\boldsymbol{\beta}_r$, the overall parameter vector $\boldsymbol{\gamma}' = ( \boldsymbol{\alpha}',\boldsymbol{\beta}')$ is partitioned as $(\boldsymbol{\beta}'_r,\boldsymbol{\phi}')$, where $\boldsymbol{\phi}$ is a vector with the remaining $(p-r)$ parameters. The unconstrained MLE is now denoted as $(\hat{\boldsymbol{\beta}'}_{r},\hat{\boldsymbol{\phi}'})$ accordingly, and the
constrained MLE is $(\hat{\boldsymbol{\beta}'}_{c,r},\hat{\boldsymbol{\phi}'}_c)$. Unconstrained CRs have the form
\begin{align}
\text{UCR}=\Big{\{} &\boldsymbol{\beta}_{0r} : 2[\ell(\hat{\boldsymbol{\beta}}_r,\hat{\boldsymbol{\phi}}) - \ell(\boldsymbol{\beta}_{0r},\tilde{\boldsymbol{\phi}})] \leq  \chi^2_{r;1-\alpha}, \boldsymbol{\beta}_{0r} \in U_{U} \Big{\}},  \label{eq:CRMT_CIupdatedMLEsubset3}
\end{align}
where $\tilde{\boldsymbol{\phi}}$ is defined by
\begin{align}
\ell(\boldsymbol{\beta}_{0r},\tilde{\boldsymbol{\phi}}) = \max_{\boldsymbol{\phi} \in U_{U}} \ell(\boldsymbol{\beta}_{0r},\boldsymbol{\phi}). \label{eq:loglikCR1}
\end{align}
Using the UCR, considering the analysis of parameters associated to a single ordinal predictor of a model with potentially more than one predictor, the following cases can occur, which are illustrated in Figure \ref{fig:UCRCases}: 
\begin{enumerate}
\item UMLE and CMLE are the same, and all parameter vectors in the CR indicate the same monotonicity direction. 
\item UMLE and CMLE are the same; the CR contains parameter vectors indicating the same monotonicity direction as the MLE, but also parameter vectors violating monotonicity.
\item UMLE and CMLE are the same; the CR contains parameter vectors indicating the same monotonicity direction as the MLE, but also parameter vectors indicating the opposite monotonicity direction, in which case there will normally also be  parameters violating monotonicity in the CR.
\item UMLE and CMLE are not the same; the CR contains parameter vectors indicating the same monotonicity direction as the CMLE on top of parameters vectors that are non-monotonic, or that potentially also indicate another monotonicity direction.
\item UMLE and CMLE are not the same, and the CR does not contain any parameter vector fulfilling the monotonicity constraints.
\item In case that there are further variables in the model, there is a further possibility, namely that the UMLE fulfills monotonicity constraints, but due to enforcement of constraints for another variable that is potentially dependent on the variable in question, UMLE and CMLE differ on that variable as well. In that case it is possible that all parameter vectors in the CR indicate the same monotonicity direction, but the other cases listed above can in principle also occur. 
\end{enumerate}

Only the first case indicates a behavior in line with the asymptotic theory. In the other cases the validity of the asymptotic CR for the given finite sample can be doubted. Assuming that monotonicity in fact holds, the other cases indicate that the true parameter is so close to the boundary of the constrained parameter space that for the given sample size one can expect an asymmetric distribution of the CMLE. In the third and potentially also the fourth case, the distribution of the CMLE may be bimodal, as the true parameter may be so close to the boundary between monotonicity directions that they may ``switch''; for datasets generated by such parameters, the UMLE may fall within both monotonicity direction domains and also may violate monotonicity, in which case the CMLE may be in either possible domain.    

One could use techniques such as bootstrap to find constrained finite sample CRs, but here we explore what can be done based on the asymptotic CRs. Section \ref{AICM_sec:CRsAndCovProb} will then explore these CRs empirically.   

The easiest way to obtain a constrained CR from an unconstrained one is to just remove all parameter vectors that do not satisfy monotonicity constraints:
\begin{align}
\text{UCCR}=\Big{\{} &\boldsymbol{\beta}_{0r} : 2[\ell(\hat{\boldsymbol{\beta}}_r,\hat{\boldsymbol{\phi}}) - \ell(\boldsymbol{\beta}_{0r},\tilde{\boldsymbol{\phi}})] \leq  \chi^2_{r;1-\alpha}, \boldsymbol{\beta}_{0r} \in U_{C} \Big{\}}.  \label{eq:CRMT_CCIupdatedMLEsubset3}
\end{align}
This may be seen as legitimate, because if the confidence level is (approximately) valid in the unconstrained case, this will also hold for the constrained case, as the constrained parameter set is a subset of the unconstrained one. This may however result in very small CRs that suggest more precision than the given sample size can provide. In case five above, the CR will even be empty. This is not incompatible with the general theory of CRs, but can be seen as undesirable.

An alternative is to center the CR at the CMLE rather than the UMLE (using the unconstrained theory otherwise), which guarantees it to be nonempty. We take into account potential constraints on other parameters by using $\tilde{\boldsymbol{\phi}}_c$ instead of $\tilde{\boldsymbol{\phi}}$, where $\tilde{\boldsymbol{\phi}}_c$ is the vector of constrained maximum likelihood estimators as a function of the value of $\boldsymbol{\beta}_{0r}$, i.e., defined according to \eqref{eq:loglikCR1}, but requiring $\boldsymbol{\phi} \in U_{C}$ rather than $U_{U}$. Therefore, we define
\begin{align}
\text{CCR}=\Big{\{} & \boldsymbol{\beta}_{0r} : 2[\ell(\hat{\boldsymbol{\beta}}_{c,r},\hat{\boldsymbol{\phi}}_{c}) - \ell(\boldsymbol{\beta}_{0r},\tilde{\boldsymbol{\phi}}_c)] \leq \chi^2_{r;1-\alpha} , \boldsymbol{\beta}_{0r} \in U_{C} \Big{\}}.  \label{eq:CRMT_CIupdatedMLEsubset2}
\end{align}

Asymptotically, at least in the interior of the parameter space, UMLE and CMLE are equal, so the UCR, UCCR, and CCR are asymptotically equivalent. On the boundary, the CCR is not even asymptotically 
guaranteed to have the nominal confidence level. Using $\tilde{\boldsymbol{\phi}}_c$ instead of $\tilde{\boldsymbol{\phi}}$ can be expected to improve the power but may incur anticonservativity (\cite{SelLia87,KopSin11}), which in the definition of the ACR below based on the CCR is not an issue. The UCCR with $\tilde{\boldsymbol{\phi}}$ is asymptotically trivially conservative but not with $\tilde{\boldsymbol{\phi}}_c$. For the CCR this would not even necessarily hold with  $\tilde{\boldsymbol{\phi}}$.
Where standard constrained inference results apply (see Chapter 4 of \cite{SilSen05}) and the constrained loglikelihood ratio follows a mixture of $\chi^2$-distributions, using the CMLE and the plain $\chi^2$-distribution will be conservative. In the given situation, this cannot be taken from granted.

A third option would be to use the aggregated CR: $$\text{ACR}=\text{UCCR}\cup\text{CCR}.$$ This guarantees a nonempty CR that is asymptotically conservative.

\subsection{Tests}\label{AICM_sec:tests}
Three kinds of null hypotheses may be of special interest for the parameters associated to an ordinal predictor $s\in\{1,\ldots,t\}$:
\begin{enumerate}
\item $H_0:\ \boldsymbol{\beta}_s=\boldsymbol{0}$; no impact of variable $s$.
\item $H_0:$ either $0 \leq \beta_{s,2}\leq\cdots\leq\beta_{s,p_s},$ or
$0 \geq \beta_{s,2}\geq\cdots\geq\beta_{s,p_s}$; variable $s$ fulfills monotonicity constraints.
\item $H_0:\ 0 \leq \beta_{s,2}\leq\cdots\leq\beta_{s,p_s}$; the impact of variable $s$ is isotonic (or, analogously, antitonic); a specific monotonicity direction obtains. 
\end{enumerate}
Only the first $H_0$ is a linear hypothesis of the type \eqref{eq:linearhyp}. This can be tested by a standard $\chi^2$ test using $R_n$, but there is a subtlety. $\hat{\boldsymbol{\gamma}}$ in \eqref{eq:likrat}, which defines $R_n$, may be taken as the UMLE, or the CMLE (or even DMLE or PMLE), respectively, in case these are different. The choice between these relies on how confident the user is to impose a monotonicity assumption, even in case that the UMLE indicates against it. 

The second $H_0$ implies that monotonicity is not assumed for the alternative. An asymptotically valid test can be defined by rejecting $H_0$ in case that no parameter vector in the UCR fulfills monotonicity constraints. In fact, it only needs to be checked whether the CMLE is in the UCR (equivalent to the UCCR not being empty), because the CMLE maximizes the likelihood among the parameter vectors that fulfill the constraints. Therefore, if the CMLE  does not fulfill $2[\ell(\hat{\boldsymbol{\beta}}_r,\hat{\boldsymbol{\phi}}) - \ell(\boldsymbol{\beta}_{0r},\tilde{\boldsymbol{\phi}})] \leq  \chi^2_{r;1-\alpha}$ with $\boldsymbol{\beta}_{0r}=\hat{\boldsymbol{\beta}}_{c,r}$, no element $\boldsymbol{\beta}_{0r}\in U_{C}$ can. The correspondence between CRs and tests is used here to define these tests, but  $p$ fulfilling  
\begin{equation}\label{epvalue}
2[\ell(\hat{\boldsymbol{\beta}}_r,\hat{\boldsymbol{\phi}}) - \ell(\boldsymbol{\beta}_{0r},\tilde{\boldsymbol{\phi}})] =  \chi^2_{r;1-p}
\end{equation}
for $\boldsymbol{\beta}_{0r}=\hat{\boldsymbol{\beta}}_{c,r}$
does not define a valid p-value, because asymptotically, under $H_0$, UMLE and CMLE are the same, so that a p-value defined in this way will converge to 1 under $H_0$.
 The opposite $H_0$, namely whether variable $s$ does {\it not} fulfill monotonicity constraints, can be tested in an analogous way by checking whether non-monotonic parameters are in the UCR.

The third $H_0$ can be tested by asking whether the corresponding PMLE (or DMLE) constraining the monotonicity direction of interest accordingly is in a suitable CR; if not, the $H_0$ is rejected.
Once more, it depends on what assumptions the user is willing to make otherwise, whether the UCCR (or, here equivalently, UCR), CCR, or ACR should be used. As long as the true parameter is in the interior of the constrained space, asymptotically these are all the same, as is the $\chi^2_r$-distribution used to define the CR.   

\section{Simulations: Finite sample performance} \label{AICM_sec:CRsAndCovProb}
\subsection{Coverage probabilities of confidence regions} \label{AICM_sec:CRsAndCovProbx}

Coverage probabilities (CPs) of the UCCR, CCR, and ACR, have been simulated and compared under different scenarios in order to see whether these asymptotically motivated CRs perform well on finite samples; see \cite{morris2019using} for some general considerations regarding such experiments. 

\begin{figure}
	\centering
	\includegraphics[width=0.8\textwidth]{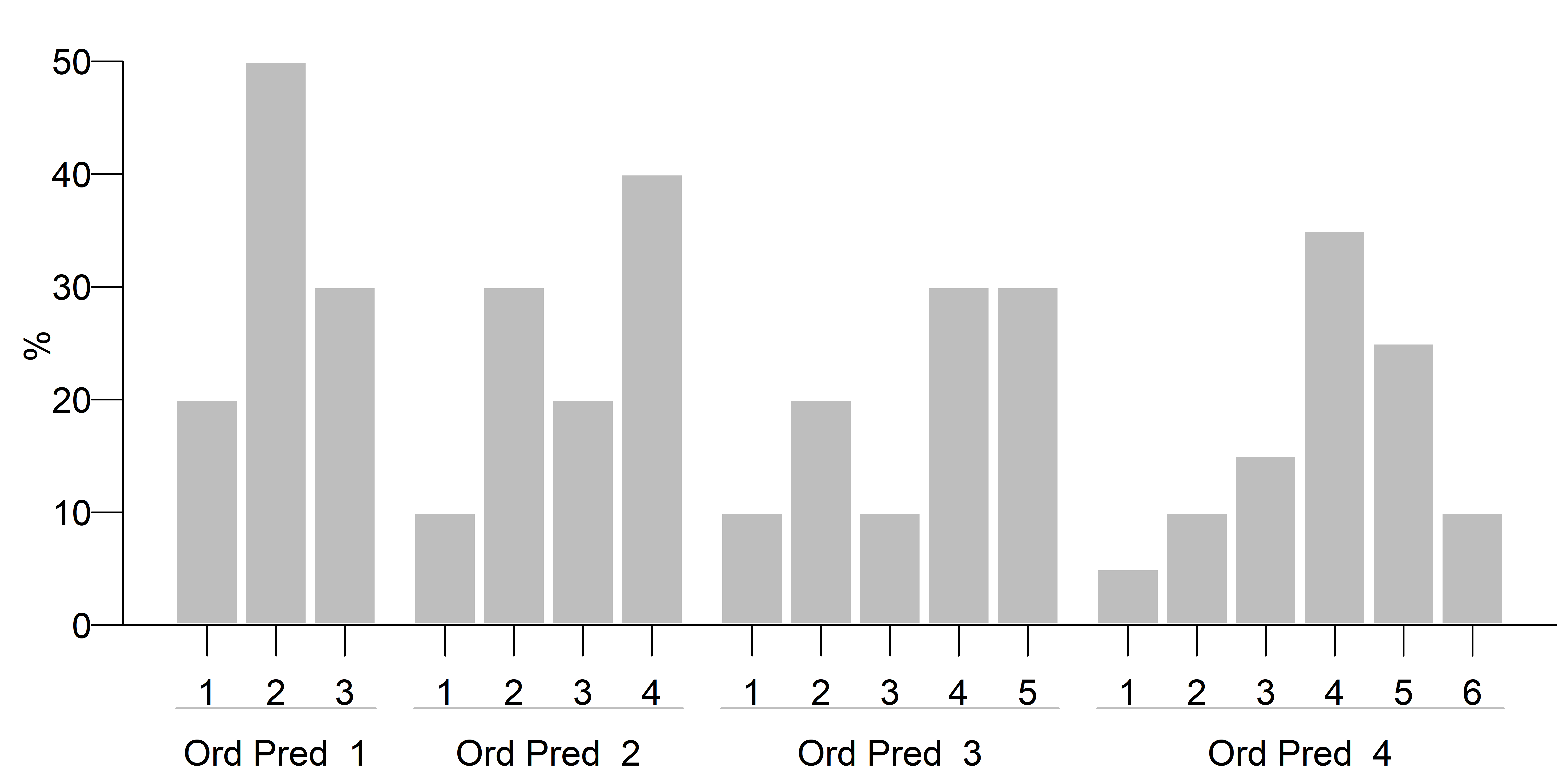}
	\caption[Distribution of ordinal categories for each simulated ordinal predictor (OP).]{Distribution of ordinal categories for each simulated ordinal predictor (OP). OPs 1 to 4 have 3, 4, 5 and 6 categories correspondingly.}
	\label{fig:OrdPredDistributionsExample1}
\end{figure}

Consider model \eqref{eq:Model_eq} with four ordinal predictors with 3, 4, 5, and 6 ordered categories each, one categorical (non-ordinal) predictor with 5 categories and one interval-scaled predictor. Each of the four ordinal predictors ($s=1,\ldots,4$) is represented in the model by dummy variables denoted as $x_{i,s,h_s}$ for observation $i$, with $h_s=2,\ldots,p_s$ and where $p_1=3$, $p_2=4$, $p_3=5$, and $p_4=6$; the nominal predictor is denoted as $x_{i,5,h_5}$ with $h_5=2,\ldots,5$; and the interval-scaled predictor as $x_{i,1}$. The first category of the categorical variables is considered as the baseline. Thus, the simulated model is 
\begin{align}
\text{logit}[P( z_{i}\leq j | \mathbf{x}_i)] =& \alpha_j+\sum_{h_1=2}^{3}\beta_{1,h_1}x_{i,1,h_1} +\sum_{h_2=2}^{4}\beta_{2,h_2}x_{i,2,h_2} + \sum_{h_3=2}^{5}\beta_{3,h_3}x_{i,3,h_3} \nonumber \\
&+\sum_{h_4=2}^{6}\beta_{4,h_4}x_{i,4,h_4}+\sum_{h_5=2}^{5}\beta_{5,h_5}x_{i,5,h_5}+\beta_{1}x_{i,1},\label{eq:Model_eq_simulCovProb}
\end{align}
where the number of categories of the ordinal response is $k=4$, i.e., $j=1,2,3$. This model was fitted for 500 data sets that were simulated using the following true parameters: for the intercepts $\alpha_1=-2$, $\alpha_2=2$, and $\alpha_3=5.5$; for the non-ordinal categorical predictor $\boldsymbol{\beta}_5'=(0.7,1.4,-0.3,-1.2)$; and for the interval-scaled predictor $\beta_1=0.3$. The values of the ordinal predictors were drawn from the population distributions represented in Figure \ref{fig:OrdPredDistributionsExample1}.
The simulated values for the non-ordinal categorical predictor were drawn from the population distribution defined by probabilities 0.2, 0.2, 0.3, 0.1, 0.2 for categories 1, 2, 3, 4, and 5, respectively. The interval-scaled covariate $x_1$ was randomly generated from $\mathcal{N}(1,4)$.

\begin{figure}
	\centering
	\includegraphics[scale=0.7]{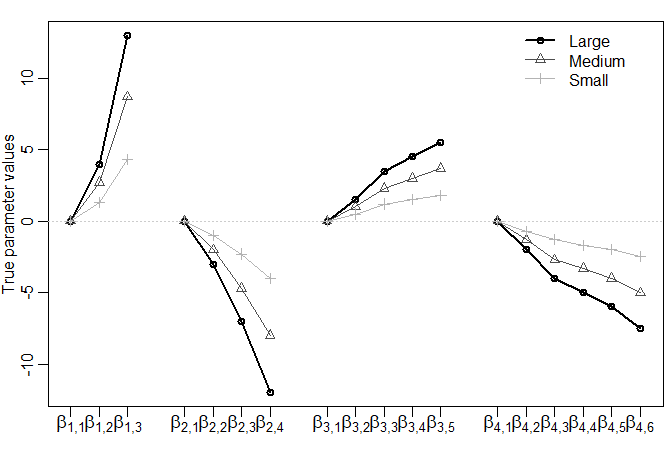}
	\caption[True parameter values for the simulation of coverage probabilities.]{True parameter values for the simulation of coverage probabilities. Different line styles represent different distances between the parameter values associated with adjacent ordinal categories: large, medium, small.}
	\label{fig:CIs_UMLEvsCMLE_Datav04trueParam_plot}
\end{figure}

The current simulation design offers 12 different scenarios defined by two factors: (i) distances between parameter values of adjacent ordinal categories and (ii) sample sizes. The true parameter vectors of the ordinal predictors were chosen to represent three different levels (large, medium, and small) of distances between the parameter values of their adjacent ordinal categories as shown in Figure \ref{fig:CIs_UMLEvsCMLE_Datav04trueParam_plot}. These will be referred to as ``monotonicity degrees''. Four different sample sizes were considered: $n=50$, 100, 500, and 1,000.

Table \ref{tab:CoverageProb1} shows the results for the three CRs defined in the previous section. The nominal confidence level was 95\%. All parameters were assessed simultaneously ($r=p$). The ``Total'' row shows the final coverage percentages. Coverage percentages are also differentiated according to whether the UMLE and CMLE are the same or not. Note that simulated coverage probabilities below 93.4\% (467/500) are significantly smaller than the confidence level of 95\%. 

\begin{table*}
\scriptsize
  \centering
\setlength{\tabcolsep}{3.8pt}
\caption{Coverage percentages (\%) for different sample sizes, definitions of CRs, and according to whether the UMLE and CMLE are the same or not. The row ``Same MLE'' has the number of cases (out of 500) in brackets for which UMLE and CMLE were the same.}
\begin{tabular}{l|ccc|ccc|ccc|ccc}
\hline\noalign{\smallskip}
$n$ & \multicolumn{3}{c|}{50} & \multicolumn{3}{c|}{100} & \multicolumn{3}{c|}{500} & \multicolumn{3}{c}{1000}  \\
CR & UCCR & CCR & ACR & UCCR & CCR & ACR & UCCR & CCR & ACR & UCCR & CCR & ACR \\
\noalign{\smallskip}\hline\noalign{\bigskip}

\multicolumn{13}{l}{\textbf{Small distances between parameter values of adjacent ordinal categories}} \\
Same MLE  & \multicolumn{3}{c|}{-   (0 cases)} & \multicolumn{3}{c|}{85.7 (7 cases)} & \multicolumn{3}{c|}{96.6 (174 cases)} & \multicolumn{3}{c}{98.4 (376 cases)} \\
Different MLE & 90.0 & 83.2 & 96.2 & 90.7 & 94.5 & 96.6 & 92.3 & 96.0 & 96.0 & 94.4 & 95.2 & 95.2 \\
\textbf{Total} & \textbf{90.0} & \textbf{83.2} & \textbf{96.2} & \textbf{90.6} & \textbf{94.4} & \textbf{96.4} & \textbf{93.8} & \textbf{96.2} & \textbf{96.2} & \textbf{97.4} & \textbf{97.6} & \textbf{97.6} \\
\noalign{\smallskip}\hline\noalign{\smallskip}

\multicolumn{13}{l}{\textbf{Medium distances between parameter values of adjacent ordinal categories}} \\
Same MLE  & \multicolumn{3}{c|}{100 (6 cases)} & \multicolumn{3}{c|}{100 (48 cases)} & \multicolumn{3}{c|}{94.6 (442 cases)} & \multicolumn{3}{c}{93.7 (495 cases)} \\
Different MLE & 94.9 & 88.5 & 98.6 & 90.5 & 94.9 & 95.6 & 82.8 & 82.8 & 82.8 & 60.0 & 60.0 & 60.0 \\
\textbf{Total} & \textbf{95.0} & \textbf{88.6} & \textbf{98.6} & \textbf{91.4} & \textbf{95.4} & \textbf{96.0} & \textbf{93.2} & \textbf{93.2} & \textbf{93.2} & \textbf{93.4} & \textbf{93.4} & \textbf{93.4} \\
\noalign{\smallskip}\hline\noalign{\smallskip}

\multicolumn{13}{l}{\textbf{Large distances between parameter values of adjacent ordinal categories}} \\
Same MLE  & \multicolumn{3}{c|}{100 (18 cases)} & \multicolumn{3}{c|}{100 (95 cases)} & \multicolumn{3}{c|}{95.2 (482 cases)} & \multicolumn{3}{c}{95.4 (500 cases)} \\
Different MLE & 97.7 & 85.7 & 99.4 & 94.3 & 95.1 & 96.8 & 77.8 & 77.8 & 77.8 & - & - & - \\
\textbf{Total} & \textbf{97.8} & \textbf{86.2} & \textbf{99.4} & \textbf{95.4} & \textbf{96.0} & \textbf{97.4} & \textbf{94.6} & \textbf{94.6} & \textbf{94.6} & \textbf{95.4} & \textbf{95.4} & \textbf{95.4} \\
\noalign{\smallskip}\hline\noalign{\smallskip}
\end{tabular}
\label{tab:CoverageProb1}
\end{table*}
% Conf Reg Constrained POCLM\12 ConfRegPOCLM_v11\R Files\Table 1\SimulationsCIsv05.xlsx

The key results are:
\begin{itemize}
\item For $n=500$ and $n=1000$ most overall coverage probabilities are close to 95\%; for ``Medium distances'' they are borderline significantly too low, caused by situations when UMLE and CMLE are different (11.6\% of the cases). Except for ``Small distances'' and $n=500$, the UMLE and CMLE are equal in most cases, as suggested by the asymptotics.
\item For low $n$, coverage probabilities of both UCCR and CCR can be substantially below 95\%, whereas the ACR is conservative. 
\item With higher monotonicity degree, coverage probabilities tend to be larger, and UMLE and CMLE tend to be the same more often.
\item For $n=50$, the UCCR has higher coverage probabilities than the CCR, whereas for $n=100$ the opposite is true. We believe that the reason is that for $n=50$ the estimators still have such a large variability that imposing the constraints may occasionally make matters worse by enforcing the wrong monotonicity direction, whereas for $n=100$ constraining starts to help.  
\item In reality researchers don't know the degree of monotonicity, and for $n=50$ only the ACR ensures good coverage for all degrees. For $n=100$ and above, the CCR achieves satisfactory results, being less conservative than the ACR.  
\end{itemize}
A further simulation addresses a case in which there is an OP with parameters at the border of the constrained parameter space $U_{C}$. Model \eqref{eq:Model_eq} with two ordinal predictors is simulated: 
\begin{align}
\text{logit}[P( z_{i}\leq j | \mathbf{x}_i)] =& \alpha_j+\sum_{h_1=2}^{3}\beta_{1,h_1}x_{i,1,h_1} +\sum_{h_2=2}^{4}\beta_{2,h_2}x_{i,2,h_2}.\label{eq:Model_eq_simulCovProbBorder}
\end{align}
OP1 has $\beta_{1,1}=\beta_{1,2}=\beta_{1,3}=0$, and OP2 has four categories using the different degrees of monotonicity from the previous setting (small, medium and large). The number of categories of the ordinal response and the remaining characteristics of the simulation setting are the same as in the first simulation above. 
The results are shown in Table \ref{tab:CoverageProb2}.

\begin{table*}
\scriptsize
  \centering
\setlength{\tabcolsep}{3.8pt}
\caption{Simulation with two OPs, where OP1 
has $\beta_{1,1}=\beta_{1,2}=\beta_{1,3}=0$.
Coverage percentages for different sample sizes, definitions of CRs, and according to whether the UMLE and CMLE are the same or not. The row ``Same MLE'' has the number of cases (out of 500) in brackets for which UMLE and CMLE were the same.}
\begin{tabular}{l|ccc|ccc|ccc|ccc}
\hline\noalign{\smallskip}
$n$ & \multicolumn{3}{c|}{50} & \multicolumn{3}{c|}{100} & \multicolumn{3}{c|}{500} & \multicolumn{3}{c}{1000}  \\
CR & UCCR & CCR & ACR & UCCR & CCR & ACR & UCCR & CCR & ACR & UCCR & CCR & ACR \\
\noalign{\smallskip}\hline\noalign{\bigskip}

\multicolumn{13}{l}{\textbf{Small distances between parameter values of adjacent ordinal categories}} \\
Same MLE  & \multicolumn{3}{c|}{97.8 (91 cases)} & \multicolumn{3}{c|}{96.0 (151 cases)} & \multicolumn{3}{c|}{96.9 (195 cases)} & \multicolumn{3}{c}{96.8 (189 cases)} \\
Different MLE & 93.6 & 93.6 & 94.1 & 91.4 & 94.6 & 94.6 & 95.4 & 96.4 & 96.4 & 93.6 & 95.8 & 95.8 \\
\textbf{Total} & \textbf{94.4} & \textbf{94.4} & \textbf{94.8} & \textbf{92.8} & \textbf{95.0} & \textbf{95.0} & \textbf{96.0} & \textbf{96.6} & \textbf{96.6} & \textbf{94.8} & \textbf{96.2} & \textbf{96.2} \\
\noalign{\smallskip}\hline\noalign{\smallskip}

\multicolumn{13}{l}{\textbf{Medium distances between parameter values of adjacent ordinal categories}} \\
Same MLE  & \multicolumn{3}{c|}{97.9 (145 cases)} & \multicolumn{3}{c|}{97.1 (172 cases)} & \multicolumn{3}{c|}{96.9 (194 cases)} & \multicolumn{3}{c}{95.3 (191 cases)} \\
Different MLE & 93.5 & 94.9 & 94.9 & 97.0 & 97.6 & 97.6 & 94.8 & 95.8 & 95.8 & 93.2 & 94.5 & 94.5 \\
\textbf{Total} & \textbf{94.8} & \textbf{95.8} & \textbf{95.8} & \textbf{97.0} & \textbf{97.4} & \textbf{97.4} & \textbf{95.6} & \textbf{96.2} & \textbf{96.2} & \textbf{94.0} & \textbf{94.8} & \textbf{94.8} \\
\noalign{\smallskip}\hline\noalign{\smallskip}

\multicolumn{13}{l}{\textbf{Large distances between parameter values of adjacent ordinal categories}} \\
Same MLE  & \multicolumn{3}{c|}{98.3 (172 cases)} & \multicolumn{3}{c|}{97.3 (182 cases)} & \multicolumn{3}{c|}{95.9 (197 cases)} & \multicolumn{3}{c}{95.3 (191 cases)} \\
Different MLE & 96.3 & 98.5 & 98.5 & 98.1 & 98.4 & 98.4 & 98.0 & 98.3 & 98.3 & 95.5 & 96.4 & 96.4 \\
\textbf{Total} & \textbf{97.0} & \textbf{98.4} & \textbf{98.4} & \textbf{97.8} & \textbf{98.0} & \textbf{98.0} & \textbf{97.2} & \textbf{97.4} & \textbf{97.4} & \textbf{95.4} & \textbf{96.0} & \textbf{96.0} \\
\noalign{\smallskip}\hline\noalign{\smallskip}
\end{tabular}
\label{tab:CoverageProb2}
\end{table*}

Different from the first simulation, for all $n$ and monotonicity degrees, both UMLE$=$CMLE and UMLE$\neq$CMLE happen with substantial probability not apparently converging to zero. This was to be expected, as in any neighbourhood of the true parameters there are both monotonic and non-monotonic parameters. The argument for the asymptotic validity of the UCCR does not hold here. Nevertheless, except for one particular combination of factors ($n=100$, ``Small distances'', and UCCR), all overall coverage probabilities are close to 95\%, none is significantly smaller, and only for large distances between the parameter values of OP2 and low $n$ there is a tendency of conservativity.

Certain results on constrained inference (e.g., \cite{GoHoMo82}) suggest that 
with the true parameter on the border of the parameter space, the distribution of the loglikelihood ratio statistic is a mixture of $\chi^2$-distributions.
In that case, its quantiles could be bounded from above by a 
$0.5\chi^2_{r}+0.5\chi^2_{r-1}$-mixture. Although known theory does not directly apply to our setup with more than one potentially constrained OP (see the Introduction), we were interested in checking whether this bound could also be valid here. To this end, for the setups discussed above, we simulated coverage probabilities using $1-\alpha$-quantiles from this mixture instead of the plain $\chi^2_{r;1-\alpha}$ used for the results in Tables \ref{tab:CoverageProb1} and \ref{tab:CoverageProb2}, for CCR and ACR (although there is no theoretical basis for doing this with ACR). The mixture quantiles are a bit smaller, and therefore the resulting overall coverage probabilities (results not shown) are slightly lower. The difference is quite small. The mixture helps a bit with the too conservative cases in Table \ref{tab:CoverageProb2}, but makes matters worse in those situations where coverage probabilities were already too low in Table \ref{tab:CoverageProb1}. In reality it is not known whether the true parameter is on the boundary, and therefore our results overall do not indicate that mixtures of $\chi^2$-distributions improve on the plain $\chi^2_r$-distribution.

\subsection{Finite sample rejection probabilities of tests} \label{AICM_sec:simulTests}

Consider model \eqref{eq:Model_eq_simulCovProbBorder} with settings and parameters values as used for the simulation presented in Table \ref{tab:CoverageProb2}. Rejection probabilities based on 500 replicates were calculated testing the three proposed hypotheses defined in Section \ref{AICM_sec:tests}, which in Tables \ref{tab:rejectionProbSmall} and \ref{tab:rejectionProbLarge} are referred to as ``first'', ``second'' and ``third'' $H_0$. We only show results for ``small'' (Table \ref{tab:rejectionProbSmall}) and ``large'' (Table \ref{tab:rejectionProbLarge}) distances between parameter values of OP2, as all results for ``medium'' are between these and do not add interesting insight. 
Each one of these tables shows the rejection probabilities for different ordinal predictors, hypotheses, confidence regions and sample sizes. 

Comparisons between Table \ref{tab:CoverageProb2} of Section \ref{AICM_sec:CRsAndCovProbx} and those of the current section are not straightforward, because coverage probabilities in Section \ref{AICM_sec:CRsAndCovProbx} were computed using confidence regions for the whole parameter vector ($r=p$), whereas rejection probabilities were computed using only two degrees of freedom when testing OP1 and three when testing OP2. Therefore, the quantiles $\chi^2_{r;1-\alpha}$ in \eqref{eq:CRMT_CCIupdatedMLEsubset3} for UCCR and in \eqref{eq:CRMT_CIupdatedMLEsubset2} for CCR differ between the sections.

Results are only shown for CCR and UCCR; results for the ACR are largely identical to those for the CCR, because parameters in UCCR$\setminus$CCR are not normally in the $H_0$.

For interpreting the results in Tables \ref{tab:rejectionProbSmall} and \ref{tab:rejectionProbLarge} keep in mind that out of the eight tests (``First'', ``Second'', ``Third'' in two directions, for both OP1 and OP2), OP1 fulfills all null hypotheses, and for OP2 two null hypotheses are violated, namely ``First (all zero)'' and ``Third (isotonic)''. The power for all tests of these two null hypotheses is 100\% for OP2. Most of the rejection probabilities for the true null hypotheses are smaller than 5\%; none is  significantly larger. For any monotonicity degree, rejection probabilities for ``Second'' and ``Third (antitonic)'' of OP2 and ``Second'' of OP1 are zero or close to it; for ``OP1: First'' and ``OP1: Third'' they are still conservative but not as extreme, less strongly so based on the CCR. Only the rejection probabilities based on the UCCR for ``OP1: First'' seem to stabilise around the nominal 5\%.

These tests do not seem to reject too easily. Loss of power due to conservativity may be a worry, although this is not confirmed for the the null hypotheses that are violated. Regarding OP2, as its monotonicity is in some distance from the border of the parameter space, conservativity is to be expected (particularly for a high degree of monotonicity), and can probably not be avoided. Regarding OP1, exact distribution theory at the border of the parameter space may make the CCR smaller, and the $H_0$ may be rejected more often in turn.  

\begin{table}[h]
\scriptsize
  \centering
\setlength{\tabcolsep}{3.8pt}
\caption{Rejection percentages (\%) based on different confidence regions for ``small'' distances between parameter values of adjacent ordinal categories.}
\begin{tabular}{l | l | l |cc|cc|cc|cc}
\hline\noalign{\smallskip}
 & & $n$ & \multicolumn{2}{c|}{50} & \multicolumn{2}{c|}{100} & \multicolumn{2}{c|}{500} & \multicolumn{2}{c}{1000}  \\
 & $H_0$ & Monotonicity Direction & UCCR & CCR &  UCCR & CCR & UCCR & CCR & UCCR & CCR \\
\noalign{\smallskip}\hline\noalign{\bigskip}
OP1 & First & None (all zero) & 1.8 & 1.4 &  2.2 & 1.6 & 2.6 & 2.0 & 5.4 & 2.8 \\
 & Second & Isotonic or Antitonic & 0.0 & 0.0 &  0.0 & 0.0 & 0.0 & 0.0 & 0.0 & 0.0 \\
 & Third & Isotonic & 1.2 & 0.8 &  1.2 & 0.8 & 1.2 & 1.0 & 2.2 & 1.2 \\
 &  & Antitonic & 0.4 & 0.4 &  1.0 & 0.8 & 1.2 & 0.8 & 2.0 & 1.6 \\
\noalign{\smallskip}\hline\noalign{\bigskip}
OP2 & First & None (all zero) & 100 & 100 &  100 & 100 & 100 & 100 & 100 & 100 \\
 & Second & Isotonic or Antitonic & 0.0 & 0.0 &  0.0 & 0.0 & 0.0 & 0.0 & 0.0 & 0.0 \\
 & Third & Isotonic & 100 & 100 &  100 & 100 & 100 & 100 & 100 & 100 \\
 &  & Antitonic & 0.0 & 0.0 &  0.0 & 0.0 & 0.0 & 0.0 & 0.0 & 0.0 \\
\noalign{\smallskip}\hline\noalign{\smallskip}
\end{tabular}
\label{tab:rejectionProbSmall}
\end{table}

\begin{table}[h]
\scriptsize
  \centering
\setlength{\tabcolsep}{3.8pt}
\caption{Rejection percentages based on different confidence regions for ``large'' distances between parameter values of adjacent ordinal categories.}
\begin{tabular}{l | l | l |cc|cc|cc|cc}
\hline\noalign{\smallskip}
 & & $n$ & \multicolumn{2}{c|}{50} & \multicolumn{2}{c|}{100} & \multicolumn{2}{c|}{500} & \multicolumn{2}{c}{1000}  \\
 & $H_0$ & Monotonicity Direction & UCCR & CCR &  UCCR & CCR & UCCR & CCR & UCCR & CCR \\
\noalign{\smallskip}\hline\noalign{\bigskip}
OP1 & First & None (all zero) & 1.8 & 1.0 &  2.2 & 0.8 & 4.4 & 3.4 & 4.4 & 2.4 \\
 & Second & Isotonic or Antitonic & 0.0 & 0.0 &  0.2 & 0.0 & 0.0 & 0.0 & 0.0 & 0.0 \\
 & Third & Isotonic & 0.8 & 0.6 &  1.0 & 0.6 & 2.0 & 1.6 & 1.6 & 1.0 \\
 &  & Antitonic & 0.6 & 0.4 &  0.8 & 0.2 & 2.2 & 2.0 & 2.4 & 1.6 \\
\noalign{\smallskip}\hline\noalign{\bigskip}
OP2 & First & None (all zero) & 100 & 100 &  100 & 100 & 100 & 100 & 100 & 100 \\
 & Second & Isotonic or Antitonic & 0.0 & 0.0 &  0.0 & 0.0 & 0.2 & 0.0 & 0.2 & 0.0 \\
 & Third & Isotonic & 100 & 100 &  100 & 100 & 100 & 100 & 100 & 100 \\
 &  & Antitonic & 0.0 & 0.0 &  0.0 & 0.0 & 0.2 & 0.0 & 0.2 & 0.0 \\
\noalign{\smallskip}\hline\noalign{\smallskip}
\end{tabular}
\label{tab:rejectionProbLarge}
\end{table}

\subsection{Comparison with inference based on one-dimensional confidence intervals}
Here we compare the proposed approach to test monotonicity with an approach based on one-dimensional confidence intervals (CIs) according to \cite{espinosa2019constrained}.

\begin{figure}
	\centering
	\includegraphics[scale=0.395]{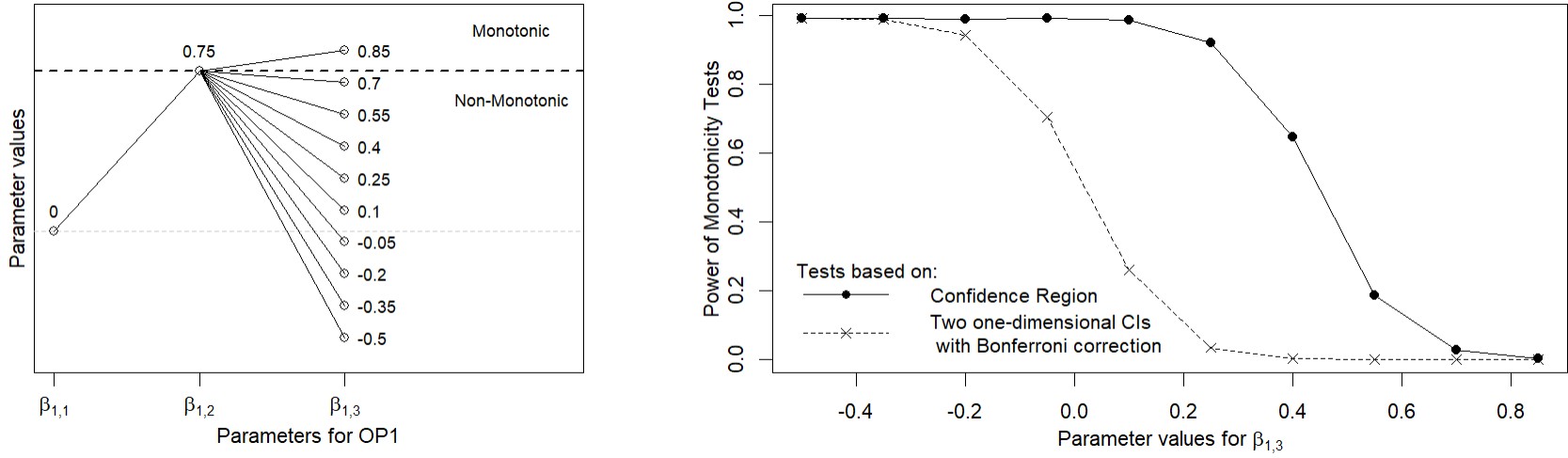}
	\caption{Left: True parameter values for the three ordinal categories of OP1, for 10 different scenarios of monotonicity levels according to the values of $\beta_{1,3}$. Right: Power of two different monotonicity tests for different levels of monotonicity of true parameter values of OP1.}
	\label{fig:simulPower}
\end{figure}

Consider model \eqref{eq:Model_eq} for an ordinal response with four categories and two OPs, $s=\{1,2\}$. OP1 has 3 categories, $h_1=\{1,2,3\}$, and OP2 has 4, $h_2=\{1,2,3,4\}$. They are represented in the model by dummy variables $x_{i,s,h_s}$. The first category of the OPs is considered baseline. Thus, the simulated model is 
\begin{align}
\text{logit}[P( z_{i}\leq j | \mathbf{x}_i)] =& \alpha_j+\sum_{h_1=2}^{3}\beta_{1,h_1}x_{i,1,h_1} +\sum_{h_2=2}^{4}\beta_{2,h_2}x_{i,2,h_2} \text{ for }j=1,2,3. \label{eq:Model_eq_simulPower}
\end{align}
This model was fitted for 500 data sets simulated using $\alpha_1=-2$, $\alpha_2=-0.5$, and $\alpha_3=0.5$. The values of OP1 were randomly drawn with probabilities 0.2, 0.5, and 0.3 for its corresponding categories 1, 2 and 3. Similarly, for OP2 the probabilities were 0.2, 0.3, 0.3, and 0.2. The parameter values for the ordinal categories of OP2 were 0, -0.3, -0.6, and -1.0. 10 different scenarios of parameters corresponding to OP1 were run. Nine of them represent different levels of non-monotonicity, going from clearly non-monotonic to almost isotonic, and the tenth case is isotonic. In the first scenario, the true parameter values for the categories of OP1 were 0, 0.75, and -0.5, being clearly non-monotonic, see plot on the left of Figure \ref{fig:simulPower}. From the second to the ninth scenario, the pattern moves closer to monotonicity by increasing the true parameter value of the third category as shown on the left of Figure \ref{fig:simulPower}.

Given that the true pattern of the parameters of OP1 is non-monotonic in nine cases, a monotonicity test should reject the null hypothesis of monotonicity. For the 500 simulated data sets, the monotonicity test proposed in Section \ref{AICM_sec:tests} was performed as well as a test that uses the Bonferroni correction on individual one-dimensional CIs as proposed and described in \cite{espinosa2019constrained}. The plot on the right of Figure \ref{fig:simulPower} shows the results for the two tests.

Unsurprisingly, the rejection rates for the null hypothesis of monotonicity decrease as the pattern of parameter values gets closer to monotonicity, see the plot on the right of Figure \ref{fig:simulPower}. This is true for both tests. The rejection rate for the test based on two individual confidence intervals decreases much earlier than the rate for the test proposed in Section \ref{AICM_sec:tests}, showing that the use of a set of one-dimensional confidence intervals for parameters loses power compared to using multivariate CRs.

\section{Application to school performance data} \label{sec:realData}

We apply the introduced methodology to data regarding the performance assessment of educational institutions in Chile. The observations are  schools that provide educational services at primary and/or secondary level. The database was built using freely accessible information\footnote{http://informacionestadistica.agenciaeducacion.cl/\#/bases}, which is published by the Education Quality Agency, an autonomous public service that interacts with the presidency of Chile through its Ministry of Education.

In Chile, the educational system considers 12 years of schooling, eight at the primary level and four at the secondary level. The unit of analysis is each school. The response variable is the performance in 2019 of each school ($z_i$ or \texttt{perf2019}), which is recorded as an ordinal variable of four ordered categories: ``Insufficient'', ``Medium-Low'', ``Medium'', and ``High''. There are two extra categories representing lack of information or low number of registered students (overall 28.3\% of the schools), which were excluded from the analysis. The analysed performance assessment results belong to 5,333 schools observed in both 2019 and 2016, specifically from their fourth year students of primary level, with national coverage. The ordinal predictors are:
\begin{itemize}
\item \texttt{perf2016}: school performance based on the 2016 assessment of fourth year students of primary level. Results are measured in the same way as the ones of the response variable \texttt{perf2019}.
\item \texttt{funding}: the way schools received their funds in 2019, recorded in three categories: ``Public'', ``Mixed'' and ``Private''. This is an ordinal predictor in the sense that it indicates the level of financial resources that each school receives per pupil from the government. The class ``Private'' receives the lowest amount, and ``Public'' receives the highest amount. The ``Private'' class has schools where the students themselves need to pay. In general, fees for ``Private'' are higher than in schools belonging to the other classes. It is usual that in Chile ``Private'' schools have more resources than others. Still, the relevant monotonicity direction of the variable may not be obvious before seeing the data, at least not without local knowledge.  
\item \texttt{regisRat}: the ratio of the number of registered students in 2019 relative to the number of 2016.
\end{itemize}
The parameter estimators from the unconstrained POCLM are given in Table \ref{tab:realDataCoef}. Estimated values for $\alpha_1$, $\alpha_2$ and $\alpha_3$, are presented as intercepts in the table. They are associated to the categories ``Insufficient'', ``Medium-Low'', ``Medium'' of \texttt{perf2019} correspondingly. The variable \texttt{perf2016} is estimated as antitonically related to $z_i$. It has negative coefficients, meaning that a better performance in 2016 is related to a better performance in 2019. This is because \eqref{eq:Model_eq} models probabilities to be {\it smaller} than certain values as dependent on $x$. The coefficients associated with the categories of \texttt{funding} are not estimated to be monotonic according to the results of the UMLE, although monotonicity is only very slightly violated. Taking the meaning of the variables into account, it can be considered highly likely that a monotonic relationship holds. ``Mixed'' funding is properly between ``Public'' and ``Private'', and reasons for potential non-monotonicity are hard to imagine. In order to improve interpretability, these parameter estimates were therefore constrained to be monotonic, and the CMLE was computed. The UMLE of \texttt{funding} is so close to being antitonic that not only is the CMLE for its parameters very close to the UMLE, also parameters of the other variables do not change by constraining \texttt{funding} to be monotonic, although theoretically this may happen. 

\begin{table}[h]
\scriptsize
  \centering
\setlength{\tabcolsep}{3.8pt}
\caption{Unconstrained (UMLE) and constrained (CMLE) estimates for school performance data.}
\begin{tabular}{ll|rr}
 &  & UMLE & CMLE \\
\noalign{\smallskip}\hline\noalign{\smallskip}
\multirow{3}{*}{Intercepts} & $\alpha_1$ & -0.62759 & -0.62759 \\
 & $\alpha_2$ & 1.83259 & 1.83260 \\
 & $\alpha_3$ & 5.87701 & 5.87701 \\
\noalign{\smallskip}\hline\noalign{\smallskip}
\multirow{3}{*}{\texttt{perf2016}} & Medium-Low & -1.23255 & -1.23255 \\
 & Medium & -3.20697 & -3.20697 \\
 & High & -5.81422 & -5.81422 \\
\noalign{\smallskip}\hline\noalign{\smallskip}
\multirow{2}{*}{\texttt{funding}} & Mixed & 0.00609 & 0.00000 \\
 & Private & -0.73117 & -0.73117 \\
\noalign{\smallskip}\hline\noalign{\smallskip}
\texttt{regisRat} &  & -0.34234 & -0.34233 \\
\noalign{\smallskip}\hline\noalign{\smallskip}
\end{tabular}
\label{tab:realDataCoef}
\end{table}

\begin{figure*}
	\centering
	\includegraphics[width=0.95\textwidth]{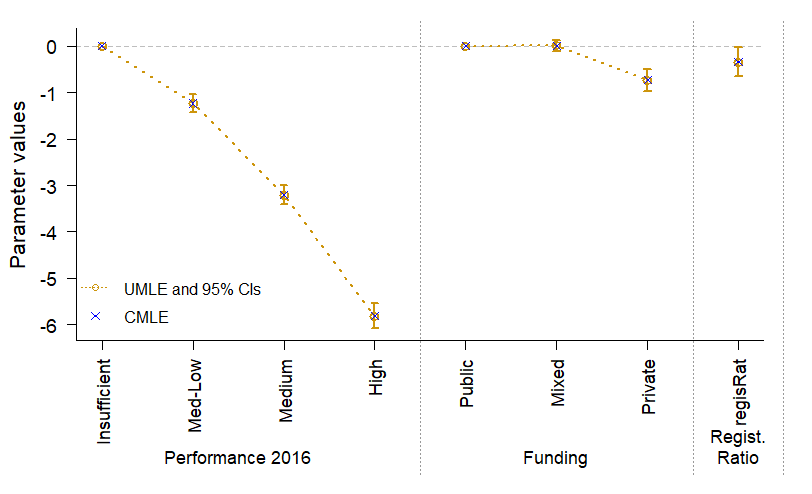}
	\caption{Coefficients estimated by UMLE and CMLE and one-dimensional 95\% confidence intervals based on the UMLE for school performance data.}
	\label{fig:school1d}
\end{figure*}

\begin{figure*}
	\centering
	\includegraphics[width=0.95\textwidth]{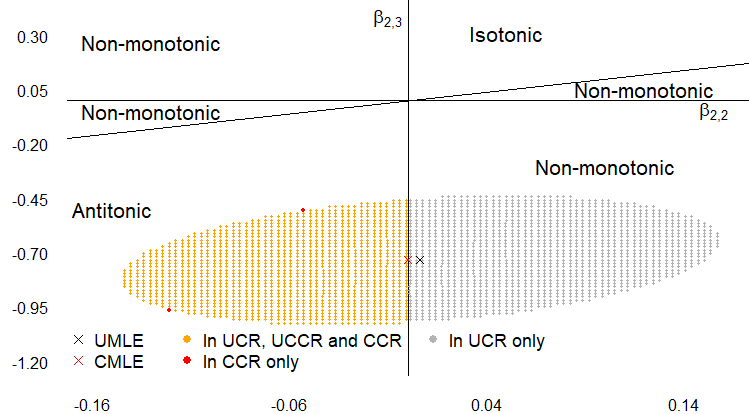}
	\caption{95\% two-dimensional confidence regions (UCR, UCCR, and CCR) for the two parameters of {\it funding}.}
	\label{fig:schoolfunding}
\end{figure*}

Figure \ref{fig:school1d} visualizes the parameter estimators with 95\% confidence intervals. 
The difference between the UMLE and CMLE is hardly visible. Figure \ref{fig:schoolfunding} shows a two-dimensional 95\% CR for the two free parameters of \texttt{funding}. 
Here the UMLE and the CMLE are visibly slightly different. 
CRs are visualized by checking whether two-dimensional projections of $\boldsymbol{\beta}_{0r}$ (for \texttt{funding} with $r=2$ this is just $\boldsymbol{\beta}_{0r}$) are in the CR for a dense grid of values.  The UCR comprises monotonic and non-monotonic parameters. The latter (colored grey) belong to the UCR, but not to the UCCR and CCR. The former (colored orange) belong to UCR, UCCR and CCR.

We highlight two differences between the UCCR and CCR. The first one is that UCCR is centered at the UMLE (even though all the non-monotonic parameters are removed) whereas the CCR is centered at the CMLE. The second one is that UCCR uses $\tilde{\boldsymbol{\phi}}$ when computing  $\ell(\boldsymbol{\beta}_{0r},\tilde{\boldsymbol{\phi}})$ as defined in \eqref{eq:CRMT_CCIupdatedMLEsubset3} whereas CCR uses $\tilde{\boldsymbol{\phi}}_c$ for $\ell(\boldsymbol{\beta}_{0r},\tilde{\boldsymbol{\phi}}_c)$ as defined in \eqref{eq:CRMT_CIupdatedMLEsubset2}. In the case of the school performance data, $\tilde{\boldsymbol{\phi}}$ and $\tilde{\boldsymbol{\phi}}_c$ are the same for every given $\boldsymbol{\beta}_{0r}$ used in the grid because the (unconstrained) parameters of \texttt{perf2016} in $\tilde{\boldsymbol{\phi}}$ turn out to show an antitonic pattern. Therefore it was not necessary to impose monotonicity (antitonic) constraints on its parameters when assessing whether $\boldsymbol{\beta}_{0r}$ was in CCR. Hence, for the school performance data, the only relevant difference between UCCR and CCR is the first one. In this application, the difference between these two CRs is not large; on the grid of parameter values for which we checked whether they are in the CR, only two (red) points belong to the CCR but not to the UCCR.

Regarding the tests in Section \ref{AICM_sec:tests}, the CR shows that the $H_0:\ \beta_{2,3}=\beta_{2,2}=0$ can be rejected; this pair of parameter values is not in the CRs.
There are monotonic parameters in the UCR, therefore the $H_0$ of monotonicity of \texttt{funding} cannot be rejected, neither can non-monotonicity be rejected. It can be rejected that  \texttt{funding} is isotonic. From a subject matter perspective, antitonicity seems likely for \texttt{funding}, and it is also well compatible with the data. Therefore it makes sense to base further inference on the CMLE.   

\begin{figure*}
	\centering
	\includegraphics[width=0.48\textwidth]{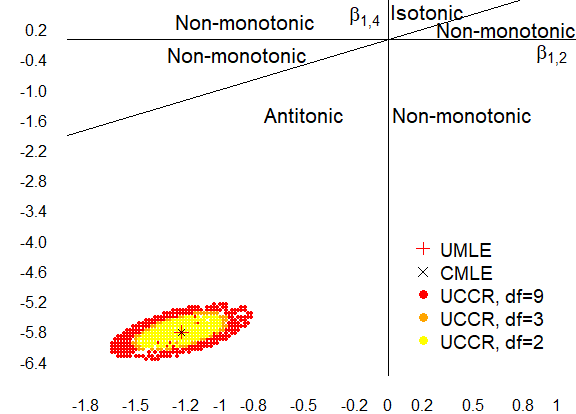}
	\includegraphics[width=0.48\textwidth]{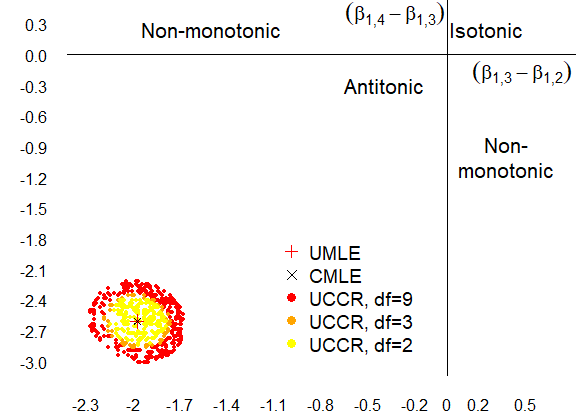}
	\caption{Left side: 95\% two-dimensional confidence regions (UCR, UCCR, and CCR are the same here) with different degrees of freedom for the $\beta_{1,4}$ vs. $\beta_{1,2}$-parameters of {\it perf2016} for school performance data. Right side: 95\% two-dimensional confidence regions (UCR, UCCR, and CCR are the same here) with different degrees of freedom for $\beta_{1,4}-\beta_{1,3}$ vs. $\beta_{1,3}-\beta_{1,2}$ of {\it perf2016}.}
	\label{fig:schoolperf}
\end{figure*}

For \texttt{perf2016} there are three free parameters, so not everything can be shown simultaneously in two dimensions. Figure \ref{fig:schoolperf} shows two selected views. On the left side, joint CRs for $\beta_{1,4}$ and $\beta_{1,2}$ are shown. Given that every $\boldsymbol{\beta}_{0r}$ in UCR is also in $U_{C}$, UCR and UCCR are the same. These two CRs are the same as CCR as well, because UMLE and CMLE here are the same, and $\ell(\boldsymbol{\beta}_{0r},\tilde{\boldsymbol{\phi}}_c)<\ell(\boldsymbol{\beta}_{0r},\tilde{\boldsymbol{\phi}})$. The plot on the left side of Figure \ref{fig:schoolperf} is of interest because the difference between $\beta_{1,4}$ and $\beta_{1,2}$ is the largest possible for this variable. This difference would need to change sign in order to revert the monotonicity direction of \texttt{perf2016}. No such parameter pair is in the CR, which suggests that the $H_0$ of isotonicity can be rejected, which can be confirmed by checking that $\hat{\boldsymbol{\beta}}_{d,r}$, the three-dimensional MLE assuming isotonicity, is not in the three-dimensional CR. An important consideration here is the number of degrees of freedom for the $\chi^2$-distribution used in the computation of the CR. If the researchers were only interested in $\beta_{1,4}$ and $\beta_{1,2}$, there should be two degrees of freedom. However, taking into account that this plot was picked because  $\beta_{1,4}$ and $\beta_{1,2}$ are the most distant parameters for this variable, choosing the degrees of freedom as 3 (number of free parameters for \texttt{perf2016}) takes into account possible variation on all three parameters for the resulting parameter pairs of $\beta_{1,4}$ and $\beta_{1,2}$. Figure \ref{fig:schoolperf} also shows the CR based on 9 degrees of freedom, which takes into account variation of all free parameters in the model. Note that the CRs are not perfectly elliptical, as the loglikelihood ratio can have irregular variations in line with asymmetries in the data, as opposed to the Wald test statistic. This is actually an advantage of the loglikelihood ratio statistic, as it reflects the information in the data more precisely.

The CR on the right side of Figure \ref{fig:schoolperf} concerns the parameter differences $\beta_{1,4}-\beta_{1,3}$ vs. $\beta_{1,3}-\beta_{1,2}$. These differences are smaller in absolute value than $\beta_{1,4}-\beta_{1,2}$, and can therefore more easily indicate potential non-monotonicity. In fact, no non-monotonic pairs of differences (let alone isotonic ones) are in the CR, so that non-monotonicity can be rejected. The same holds for the $H_0$ that \texttt{perf2016} does not have any impact. On top of these tests researchers of course may be interested  in ``effect sizes'', i.e., how far the parameter values are away from non-monotonicity or isotonicity relative to the variation of the parameter estimators, which the CRs visualize as well.

In general, for variables with more than three categories, and consequently $r>2$ free parameters, one of the most interesting two-dimensional views may be the one that shows the most extreme parameter pair (left side of Figure \ref{fig:schoolperf}). This helps to assess how far the data are from the inverse monotonicity direction. Further views of interest are one or more plots of pairs of differences that either in terms of the UMLE run counter to the dominating monotonicity direction, or are the closest to violating it (right side of Figure \ref{fig:schoolperf}; even though for this variable all plots comfortably confirm antitonicity).

\section{Conclusion} \label{sec:Conclusions}
We believe that the monotonicity assumption plays a key role in regression problems with both ordinal predictors and an ordinal response. Obviously, monotonicity cannot be taken for granted in every application, but as is the case for the linear assumption for interval scaled variables, it means that the impact of a predictor on the response works in the same way over the whole range of the scale.  

 \cite{espinosa2019constrained} already proposed an approach to decide about whether monotonicity is appropriate at all, and what monotonicity direction to choose for which variable. This was based on one-dimensional confidence intervals, not taking the potential dependence between parameters appropriately into account, resulting in weaker power than possible. Here we show that looking at higher dimensional CRs the asymptotic theory of the unconstrained POCLM applies to the model with monotonicity constraints as well, unless parameters are on the boundary of the constrained space. Finite samples can highlight issues with the asymptotic approximation, namely where UMLE and CMLE are different, and where non-monotonic parameter vectors are in the UCR. For testing monotonicity, the unconstrained theory can be used, but assuming monotonicity, CRs need to be adapted. We propose some simple ways of doing that, and some tests of standard hypotheses of interest that are easy to evaluate. Finite sample theory is not easily available, neither is asymptotic theory on the boundary, therefore the CRs can only be validated empirically. Alternatives worthy to explore are bootstrap tests and confidence regions  (\cite{hall1987bootstrap,olive2018regions}), although multivariate bootstrap CRs are often computationally expensive and rarely applied in the literature. It may also be possible if probably hard to adapt existing approaches to constrained inference (in particular potentially the approach in \cite{AMVZCaGo20,CoxShi22}) to the situation studied here. An interesting refinement of our approach, without deriving theory for parameters on the boundary, could be to consider parameters in the interior of the parameter space that approach the boundary at speed $\mathcal{O}(1/\sqrt{n})$ in Theorem 1. Such a result does currently not exist.

Generally there is much scope for exploring alternative ideas, and our approach is probably not the last word on the matter, but we hope that it can stimulate further research.

As opposed to most work on constrained inference, we focus on CRs and define tests indirectly by correspondence to CRs. Disadvantages are the lack of automatically produced p-values, and loss of power compared with potential procedures based on constrained theory for parameters on the boundary, although our simulation study did in most situations not result in too large overconservative CRs. For small samples they could even be anticonservative. But there are also advantages. There is currently much controversy regarding significance tests and p-values and their endemic abuse. A major issue with them is that researchers often focus on significance only and neglect effect sizes (\cite{WasLaz16,Stahel21}), which CRs enforce to take into account. Furthermore, basing the CRs on unconstrained theory may help in situations in which a researcher is not sure about imposing monotonicity; information is always implied to assess this assumption.

\section{Statements and Declarations} 

\subsection{Competing Interests} 

There are no competing interests.

\subsection{Funding} 
This research was partially funded by DICYT project 032262EB, Vicerrectoría de Investigación, Desarrollo e Innovación of the University of Santiago of Chile (USACH).

\subsection{Data Availability} \label{sec:DataAvailability}
Datasets for the real data application on school performance are of public access. They are available in the Studies Portal of the Agency of Education Quality, an autonomous public service that interacts with the presidency of Chile through its Ministry of Education. The link of its website is \texttt{https://informacionestadistica.agenciaeducacion.cl/\#/bases}. Two datasets were used, one of 2016 and another of 2019. They can be found using the following options: choose ``Categorías de desempeño'' for ``Proceso'', 2016 or 2019 for ``Año'', and ``Educación Básica'' for ``Grado''.

\bibliographystyle{apalike}

\end{document}